\RequirePackage{fix-cm}
\documentclass[twocolumn,epjc3]{svjour3}  
\usepackage{mathtools, cuted}
\usepackage{graphicx}
\usepackage{amsmath}
\usepackage{amssymb}
\usepackage{mathrsfs}
\usepackage[final]{pdfpages}
\newcommand{\Lagr}{\mathcal{L}}
\usepackage{latexsym}
\usepackage[numbers,sort&compress]{natbib}
\usepackage[colorlinks,citecolor=blue,urlcolor=blue,linkcolor=blue]{hyperref}
\usepackage{xcolor}
\hypersetup{
  colorlinks=true,        
  linkcolor=blue,         
  citecolor=magenta,      
}
\journalname{Eur. Phys. J. Plus, }

\begin{document}
\title{Constraining Theories of Gravity by GINGER experiment}

\author{ 
Salvatore Capozziello\thanksref{addr1,addr11, e1}
\and Carlo Altucci \thanksref{addr1}
\and Francesco Bajardi \thanksref{addr1}
\and Andrea Basti\thanksref{addr2,addr3}  
\and Nicol\`o Beverini\thanksref{addr2} 
\and Giorgio Carelli\thanksref{addr2} 
\and Donatella Ciampini\thanksref{addr2}
\and Angela D. V. Di Virgilio\thanksref{addr3}
\and Francesco  Fuso\thanksref{addr2} 
\and Umberto Giacomelli\thanksref{addr3} 
\and Enrico Maccioni\thanksref{addr2,addr3} \and Paolo Marsili\thanksref{addr2} 
\and Antonello Ortolan\thanksref{addr4}
\and Alberto Porzio \thanksref{addr5}
\and Andrea Simonelli\thanksref{addr3} 
\and Giuseppe Terreni\thanksref{addr3}
\and Raffaele Velotta\thanksref{addr1} }
\thankstext{e1}{e-mail: capozziello@na.infn.it}
\institute{
 Dipartimento di Fisica Ettore Pancini, Universit\`a di Napoli Federico II and INFN sez. di Napoli,  Complesso Univ.\ Monte Sant'Angelo, via Cintia, Napoli, Italy,\label{addr1}
 \and
 Scuola Superiore Meridionale, Largo S. Marcellino 10, I-80138 Napoli, Italy, \label{addr11}
\and
 Dipartimento di Fisica Enrico Fermi, Universit\`a di Pisa,  Largo B.~Pontecorvo 3, Pisa, Italy,\label{addr2}
 \and
 INFN Sez.~di Pisa, Largo B. Pontecorvo 3, Pisa, Italy,\label{addr3}
\and
 INFN-National Laboratories of Legnaro, viale dell'Universit\`a 2, I-35020, Legnaro (PD), Italy,\label{addr4}
 \and
 CNR-SPIN and INFN, Napoli, Complesso Univ.\ Monte Sant'Angelo, via Cintia, Napoli, Italy. \label{addr5}
}
\date{\today}
\maketitle
\begin{abstract}
The debate on  gravity theories to extend or modify General Relativity is very active today because of the issues related to ultra-violet  and infra-red behavior of Einstein's theory. In the first case, we have to address the Quantum Gravity problem.  In the latter, dark matter and dark energy, governing the large scale structure and the cosmological evolution, seem to escape from any final fundamental theory and detection.  The state of art is that, up to now, no final theory, capable of explaining gravitational interaction at any scale, has been formulated. In this perspective, many research efforts are devoted to test theories of gravity by space-based experiments. Here we propose straightforward tests by  the GINGER experiment, which, being Earth based, requires little modeling of external perturbation, allowing a thorough analysis of the systematics, crucial for experiments where sensitivity breakthrough is required. Specifically, we want to show that it is possible to constrain parameters of gravity  theories, like  scalar-tensor  or Horava-Lifshitz gravity, by considering their post-Newtonian limits matched with experimental data. In particular, we use the Lense-Thirring measurements provided by GINGER to find out  relations among the  parameters of  theories and finally compare the results with those provided by LARES and Gravity Probe-B satellites.
\end{abstract}

\hspace*{-0.5cm}\textbf{Pacs} \, {04.50.Kd; 04.80.Cc; 91.10.Op.}
\section{Introduction}
  General Relativity (GR) is one of the cornerstones of modern Physics predicting the behavior of gravitational field  from  very large astrophysical scales up to local scales. For instance, it provides corrections to the path of planets orbiting around stars, which undergo a  perihelion precession which cannot be theoretically obtained by  Newtonian gravity. Moreover, in the last few years, the gravitational waves detection \cite{Abbott:2016blz, TheLIGOScientific:2017qsa, Abbott:2017vtc, Monitor:2017mdv, Abbott:2017oio} and the confirmation of the existence of black holes \cite{Akiyama:2019cqa} further corroborated the validity of the Einstein geometric view of  gravitational interaction. The theory satisfactorily explains most of the cosmic evolution, starting  from the very early  up to the present  epoch giving rise to the so called Cosmological Standard Model \cite{Weinberg}. For these reasons, nowadays GR is the best accepted theory describing gravity. 
  
 Nevertheless, though it  fits a large amount of   observational data, it suffers many problems at very large and very small scales. Several shortcomings arise in the attempt to coherently describe the large-scale structure and cosmology. In particular, recent observations suggest that the universe is undergoing an  accelerated expansion in the late epoch. The current explanation for this issue is that the expansion should be driven by a cosmological constant $\Lambda$ acting at large scales or some form of  \emph{dark energy} which allows the evolution of cosmological constant from early to present epochs. According to the   observations,  dark energy  represents almost  70$\%$ of the whole matter-energy content of the  universe, but a final explanation of its fundamental nature is still missing.
 
  On the other hand, the  velocity of  stars and gas clouds orbiting around galaxies  led to introduce an unknown form of \emph{dark matter} \cite{Sahni:2004ai, Catena:2009mf, Navarro:1995iw, Bertone:2004pz}. Its existence was firstly inferred by the observations of the galaxy rotation curves.   Astrometric data suggest that  dark matter in the universe is of the order of 25 - 30 $\%$ of the total amount of cosmic matter-energy leaving very little room for the observed  ordinary baryonic matter.  Also in the case of dark matter, there is no final evidence, at fundamental quantum level, also if its cumulative effects are present at astrophysical scales.
  
  Dark energy and dark matter constitute the most striking  shortcomings of Cosmological Standard Model if no final signature of their existence is revealed at fundamental level, despite of the several  models  proposed to explain them. 
  
 Furthermore,   at  ultraviolet scales, GR cannot be dealt under   the standard formalism of Quantum Field Theory. The reason   is because space and time cannot be simply considered as quantum variables.  Several approaches are under scrutiny like the canonical formalism, started from the early works by Arnowitt, Deser, and Misner  \cite{Thiemann:2007zz, Arnowitt:1960es, Misner:1969ae, Arnowitt:1962hi, Bajardi:2020fxh},  the Quantum Field Theory formulation on curved spacetimes \cite{DeWitt:1975ys, Padmanabhan:2003gd, Calzetta:1986ey},   the Superstring Theory  \cite{String}, the Supergravity \cite{VanNieuwenhuizen:1981ae} and others, but none, up to now, can be considered the  final,  self-consistent formulation of Quantum Gravity \cite{Kiefer:2003ew}.
 
 However, under some assumptions,  the semiclassical  formulation (that is quantum fields formulated on curved "classical" spaces) is giving rise to physically consistent results related, for example, to black holes  \cite{Hawking:1974sw, Bekenstein:1973ur, Strominger:1996sh, Gibbons:1976ue, Banados:1992wn}, to the inflationary universe \cite{Guth:1980zm, Linde:1981mu, Linde:1983gd, Guth:1982ec, Elizalde:2007zz} and to other topics related to the behavior of gravitational field in strong regime. 
 The paradigm is that, in the semiclassical limit, the gravitational   action can be corrected with  curvature invariants, scalar fields or other geometric invariants (like torsion) in order to represent the effective gravitational interaction \cite{Birrell, Buchbinder:1992rb}.
 As a result, the Hilbert-Einstein action of gravity results extended or modified giving the possibility to fix shortcomings of GR at infra-red and ultra-violet scales.
 
 Essentially, these effective actions can be distinguished in two main categories: \emph{Extended Theories of Gravity} \cite{Capozziello:2011et, Wands:1993uu, Crisostomi:2016czh, Clifton:2011jh, Nojiri:2006ri, Bajardi:2020mdp, Bajardi:2020osh, Bajardi:2019zzs} which improve the Hilbert-Einstein  action involving  higher-order curvature invariants or scalar fields, and \emph{Alternative Theories of Gravity}, where some basic  assumptions of GR are relaxed, as   well as the torsionless connection,  the metricity, or the universal validity of   the Equivalence  Principle  \cite{Bajardi:2021tul, Cai:2015emx, Hammond:2002rm, Sotiriou:2006qn, BeltranJimenez:2019tjy, Vitagliano:2010sr, Frusciante:2015maa}. 
 
 For instance, $f(R)$ gravity is a straightforward generalization of GR  where  a generic function of the  Ricci curvature scalar $R$ is introduced relaxing the linearity in $R$. In this case, the field equations exhibit an effective energy-momentum tensor where further curvature components  can play the role of dark energy and dark matter \cite{Capozziello:2002rd,  Sotiriou:2008rp, DeFelice:2010aj, Starobinsky:2007hu, Nojiri:2006gh, Capozziello:2006dj, Cognola:2007zu}. 
In particular, it is possible to show that such a curvature stress-energy tensor can be recast in a perfect fluid form where cosmological dynamics is ruled by the form of $f(R)$ function \cite{Capozziello:2018ddp}.
 
 It is possible to show that a conformal transformation   provides the Hilbert-Einstein  action minimally coupled to a scalar field $\phi$  related to the first derivative of $f(R)$ function \cite{Capozziello:2015hra, Borowiec:2014wva, Stabile:2013eha}. This means that Extended Theories of Gravity can be recast  in equivalent forms where  GR is minimally coupled to one or more  scalar fields depending on the  involved degrees of freedom of gravitational action. 
 Scalar fields are usually introduced  to drive the cosmological inflation at   early times and to mimicking dark energy behavior at late time  of the universe expansion \cite{Capozziello:2002rd}.
 
Other important classes  of alternative theories of gravity are  the so called \emph{Teleparallel Gravity} \cite{Aldrovandi:2013wha}, \emph{Symmetric Teleparallel Gravity} \cite{Conroy:2017yln},  \emph{Horava--Lifshitz gravity} \cite{Horava:2009uw} where the Lorentz--invariance turns out to be broken at  fundamental level. In this perspective, it is important to point out that  an apparatus like GINGER, for testing the Lorentz Invariance,  is reliable  as reported  by J. Tasson and collaborators in the framework of the Extended Standard Model of Particles \cite{Jay2019}.

In general, modifying or extending the gravitational action, yields further degrees of freedom. Even the simplest extensions, like $f(R)$ gravity, lead to further gravitational waves modes \cite{Capozziello:2019klx} and to fourth--order field equations, in the metric formalism,   depending on the function $f(R)$. 

Several phenomenological selection criteria  aim to find out the form of $f(R)$ by best fitting   observations. In \cite{Capozziello:2006ph, Capozziello:2004us}, for example, the theory is selected by means of galaxy rotation curves; in \cite{Santos:2007bs},  cosmographic parameters are used in order to find the shape of action respecting the energy conditions. In \cite{Sotiriou:2008ya}, actions providing  bouncing universes are taken into account. From an astrophysical point of view, theories of gravity can be constrained according to the parameters of compact objects \cite{Astashenok} or the multimessenger astronomy combining gravitational waves and electromagnetic signals \cite{Lombriser:2015sxa}.
Finally, space-based experiments like LARES and GravityProbe B satellites can give upper bounds on gravitational parameters (see e.g. \cite{Capozziello:2014mea}).

In principle, also ground-based experiments can be considered to constrain theories of gravity at some fundamental level. The main advantages of Earth based experiments are: $i)$ that they provide  local responses, not  averaged ones; $ii)$ they can be repeated in different locations; $iii)$ they do not require external perturbation modelling, nor independent gravity maps; $iv)$  synchronization and tuning are simpler than analogous space-based experiments; $v)$ moreover,  the apparatuses  can be periodically upgraded to refine the measurements. 

In this paper,  we propose  to  constrain  theories of gravity exploiting the expected sensitivity on relativistic precessions of the GINGER (Gyroscopes IN General Relativity) experiment. In particular, we shall take into account  metric  theories, whose action includes  curvature invariants and  scalar fields.   Another theory which we are going to test  is the  Horava--Lifshitz gravity which we shall describe below. In both cases, the scalar and vector potentials coming from the weak--field limit are derived. The aim is to constrain the free parameters of gravitational potentials by GINGER experimental data. 

Specifically, GINGER  is an Earth-based experiment suitable  for measuring the Lense-Thirring and Geodesic precessions of the rotating Earth. The measurements  of the rotation rate, by means of an array of ring lasers,  provide the GR components of the gravito-magnetic field with a precision of at least 1\% \cite{Tartaglia:2016jfo, DiVirgilio:2020ior, DiVirgilio:2017fuh, Ruggiero:2015gha, Bosi:2020dgg, DiVirgilio:2021zmm}. Recent results of its prototype GINGERino, at Gran Sasso Laboratories \cite{ DiVirgilio:2021zmm}, indicate that GINGER should be able to measure  Geodesic and  Lense-Thirring effects with an uncertainty of 1 part in $10^4$ and $10^3$, respectively\footnote{In the first GINGER proposal, the conservative target was 1 part in $10^2$ of the Lense--Thirring effect, but the data analysis of the existing prototypes is showing that the sensitivity is better than expected, and a factor 10 improvement feasible.},  of their value calculated in the framework of GR.
Therefore, we can falsify or constrain the parameter space of extended/modified theories of gravity by comparing their post--Newtonian (PN) and parameterized post--Newtonian (PPN) predictions to the corresponding GINGER measurements.  
The general purpose is to demonstrate that theories of gravity can be constrained not only by astrophysical and cosmological observations or space-satellites but also by terrestrial experiments that can be, in principle, more easily tuned and controlled \cite{Gurzadyan,Gurzadyan1}. 

The paper is organized as follows: in Sect. \ref{ETG}, we overview the main aspects of extended/modified theories of gravity, paying attention to higher-order-scalar-tensor theories and Horava--Lifshitz gravity. These theories  can considered  general schemes among the alternatives to GR. In Sect. \ref{KERR}, after briefly introducing the features of Kerr spacetime, we take into account  the PN limit of the  considered theories, finding the explicit form of the Gyroscopic and the Lense--Thirring precessions. Afterwards, we use such precessions to constrain  gravity models by GINGER data. Finally, in Sect. \ref{CONCL},  we discuss the main results and the future perspectives of the approach.

\section{Examples of Extended/Modified Theories of Gravity}
\label{ETG}
As mentioned in the Introduction, modified theories of gravity aim to relax some assumption of GR, as well as that of second--order field equations or symmetric connections while extended theories retain the fundamental assumptions of GR but take into account further ingredients into the gravitational actions like curvature invariants and scalar fields. The motivations come, essentially, from addressing the behavior of gravitational interaction at ultra-violet and infra-red scales. In this section,  we are going to overview some   aspects of these theories, mainly focusing on two classes of theories, which can be considered as a sort of paradigmatic approach in this topic, with the aim  to constrain the weak-field  parameters by experimental observations. 

Let us start discussing   Extended Theories of Gravity \cite{Capozziello:2011et}. They usually extend the Hilbert--Einstein action to a function of  curvature invariants and scalar fields. The simplest extension is the so-called $f(R)$ gravity, whose action reads:
\begin{equation}
S= \int \sqrt{-g} \, f(R) \, d^4 x.
\label{FRA}
\end{equation}
The action depends on a general function of the curvature scalar  $R$. The field equations can be obtained by varying the action with respect to the metric $g_{\mu \nu}$:
\begin{eqnarray}
G_{\mu \nu} &=& \frac{1}{f_R(R)}  \left\{\frac{1}{2}g_{\mu\nu} \left[f(R)-f_R(R)R\right] \right. \nonumber
\\
&& \left. + \left[\nabla_\mu \nabla_\nu - g_{\mu \nu} \Box \right] f_R(R)\right\},
\label{FEFR}
\end{eqnarray}
where the subscript denotes the derivative with respect to $R$, $\nabla_\mu$ is the covariant derivative and $\Box$ is the d'Alembert operator $\Box = g_{\mu \nu} \nabla^\mu \nabla^\nu$. Comparing Eq. \eqref{FEFR} with vacuum Einstein field equations $G_{\mu \nu} = 0$, we notice that the RHS can be intended as an effective curvature energy--momentum tensor, namely
\begin{eqnarray}
T^{curv}_{\mu \nu} &=&  \frac{1}{f_R(R)} \left\{\frac{1}{2}g_{\mu\nu} [f(R)-f_R(R)R] \right. \nonumber
\\
&& \left. + \left[\nabla_\mu \nabla_\nu - g_{\mu \nu} \Box \right] f_R(R)\right\},
\end{eqnarray}
In this way, geometric contributions can be recast as a geometric fluid capable of describing dark energy and dark matter effects \cite{Capozziello:2018ddp}. 

Among the $f(R)$ extensions, one of the most studied is the Starobinsky model \cite{Starobinsky:1980te, Starobinsky:1982ee}, whose action consider second-order corrections to the Ricci scalar. It is
\begin{equation}
S= \frac{1}{16 \pi G_N} \int \sqrt{-g} \, (R + \alpha R^2)\, d^4 x.
\end{equation}
with $G_N$ being the Newton constant. It fits very well early--time inflation prescriptions according to the PLANCK experiment data \cite{Ade:2015rim}. 

A further extension  involving   scalar--tensor degrees of freedom with a self-interaction potential $V(\phi)$, a non-minimal coupling $F(\phi)$ and kinetic term $\omega(\phi)$, can be written as:
\begin{equation}
S = \int \sqrt{-g} \left[R F(\phi) + \omega(\phi) \nabla_\mu \phi \nabla^\mu \phi + V(\phi) \right] d^4 x.
\label{STA}
\end{equation}
It can be shown that under appropriate  conformal transformations,  the action \eqref{FRA} and the action \eqref{STA} can be recast under the same standard of an Einstein theory plus a scalar field \cite{Bajardi:2020xfj}.
Specifically, $f(R)$ gravity corresponds to  $\omega=0$ in the metric case \cite{Capozziello:2011et}.
The action \eqref{STA} can be further generalized by introducing second order curvature invariants as $R^{\mu \nu} R_{\mu \nu}$ and the scalar curvature\footnote{Introducing also the quadratic Riemann invariant $R^{\alpha\beta\mu\nu}R_{\alpha\beta\mu\nu}$ does not add further information thanks to the Gauss-Bonnet topological term ${\cal G}=R^2- 4R^{\mu\nu}R_{\mu\nu}+R^{\alpha\beta\mu\nu}R_{\alpha\beta\mu\nu}$ which fixes a relation between the 3 curvature quadratic invariants.}. It reads:
\begin{equation}
S=\int  \sqrt{-g}\left[f\left(R, R_{\alpha \beta} R^{\alpha \beta}, \phi\right)+\omega(\phi) \nabla_\alpha \phi \nabla^\alpha \phi\right] d^{4} x
\label{GENACT}
\end{equation}
and includes all the above cases. The variation of the action with respect to the metric tensor yields the field equations:
\begin{equation}
\begin{array}{l}\displaystyle f_{R} R_{\mu \nu}-\frac{f+\omega(\phi) \nabla^\alpha \phi \nabla_\alpha \phi }{2} g_{\mu \nu}-\nabla_\mu \nabla_\nu f_{R} \\ \\ +g_{\mu \nu} \square f_{R}+2 f_{Y} R_{\mu}^{\alpha} R_{\alpha \nu}  \displaystyle  -2 f_{Y}(\nabla_{\alpha} \nabla_{\nu} R^\alpha_\mu + \nabla_{\alpha} \nabla_{\mu} R^\alpha_\nu) \\ \\ +\square\left(f_{Y} R_{\mu \nu}\right)+g_{\mu \nu} \nabla_\beta \nabla_\alpha\left(f_{Y} R^{\alpha \beta}\right)+\omega(\phi) \nabla_\mu \phi \nabla_\nu \phi =0 , \end{array}
\label{FEGEN}
\end{equation}
where $\nabla$, as above,  denotes the covariant derivative with respect to the Levi-Civita connection, the subscript denotes the partial derivative and $Y$ is defined as $Y \equiv R^{\mu \nu} R_{\mu \nu}$. The dynamics of the scalar field is ruled by the Klein-Gordon equation, that is
\begin{equation}
2 \omega(\phi) \square \phi+\omega_{\phi}(\phi) \nabla_\alpha \phi \nabla^\alpha \phi -f_{\phi}=0.
\end{equation}
 Notice that both the scalar--tensor action \eqref{STA}  and the $f(R)$ action \eqref{FRA} are particular cases of  the general action \eqref{GENACT}. Constraining this theory, therefore, automatically implies constraining several Extended Theories of Gravity. In Table I, we report various Extended Theories of Gravity, the effective PN potentials, and the free parameters characterizing them.

As an example of  alternative theory to GR, we consider  the Horava-Lifshitz gravity, proposed by Horava in \cite{Horava:2009uw}. It has been formulated as  an  effective Quantum Gravity approach not requiring the Lorentz invariance at  fundamental ultra-violet scales.  This invariance,  however, emerges at large distances. It mainly aims to solve the high--energy issues suffered by GR by means of a spacetime foliation capable of reproducing the causal structure out of the quantum regime. Basic foundations and applications of this approach can be found \emph{e.g.} in \cite{Horava:2009if, Lu:2009em, Calcagni:2009ar, Charmousis:2009tc, Brandenberger:2009yt, Sotiriou:2009bx, Cai:2009pe, Panotopoulos:2020uvq, Vernieri:2019vlh, Sotiriou:2014gna}. The starting action is
\begin{eqnarray}
S = \int \sqrt{-g} N && \left\{ \frac{1}{16 \pi G_{HL}} (K^{ij} K_{ij} - \lambda K^2 ) \right. \nonumber \\ 
&& - \frac{16 \pi G_{HL} }{w^4} \big[ \nabla_i R_{jk} ( \nabla^i R^{jk} - \nabla^j R^{ik} )  \nonumber
\\
&& \qquad \qquad   \left. - \frac{1}{8} \nabla_i R \nabla^i R \big] \right\} d^3 x \, dt,  
\label{HLACT}
\end{eqnarray}
where $N^i$ and $N$ are lapse and shift functions, respectively, defined by means of the metric interval as
\begin{equation}
d s^{2}=N^{2} d t^{2}-g_{i j}\left(d x^{i}+N^{i} d t\right)\left(d x^{j}+N^{j} d t\right),
\end{equation}
while $K_{ij}$ and $K$ are\footnote{Latin indexes label the spatial coordinates.}
\begin{equation}
K_{i j}=\frac{1}{2 N}\left(\dot{g}_{i j}-\nabla_{i} N_{j}-\nabla_{j} N_{i}\right) \qquad K=g_{i j} K^{i j}.
\end{equation}
The constant $\lambda$ measures the deviation of the kinetic term from GR. The parameter $w$ is a dimensionless coupling and $G_{HL}$ becomes $G_N$ as soon as GR is recovered. In a spherically symmetric spacetime, the solution of the field equations occurs analytically and has the form \cite{Kehagias:2009is, Harko:2009qr}
\begin{equation}
g_{00} = g_{11}^{-1} = 1+w r^{2}-w r^{2} \sqrt{1+\frac{4 G_{HL} M}{w r^{3}}}.
\end{equation}
The Schwarzschild spacetime can be recovered as soon as $ 4 MG_{HL}\ll wr^3$. 

A generalized Horava-Lifshitz action is

\begin{eqnarray}
&& S = \frac{1}{16 \pi G_{HL}} \int  N \sqrt{g} \left\{ K_{i j} K^{i j}-\lambda K^{2} - 2 \Lambda+R \nonumber \right. 
\\ 
&& - 16 \pi G_{HL} \left(g_{2} R^{2}+g_{3} R_{i j} R^{i j}\right)  -(16 \pi G_{HL} )^2 \left[g_{4} R^{3} \right. \nonumber
\\
&& +g_{5} R R_{i j} R^{i j} +g_{6} R_{j}^{i} R_{k}^{j} R_{i}^{k} +g_{7} R \nabla^{2} R \nonumber
\\
&& \left. +g_{8}\left(\nabla_{i} R_{j k}\right)\left(\nabla^{i} R^{j k}\right)\right] +\phi \mathcal{G}^{i j}\left(2 K_{i j}+\nabla_{i} \nabla_{j} \phi\right) \nonumber
\\
&& \left. + \frac{\mathscr{A}}{N}\left(2 \Lambda-R\right) \right\} d t d^{3} x,
\end{eqnarray}
with $\mathcal{G}_{ij}$ being the 3D-Einstein tensor
\begin{equation}
\mathcal{G}_{i j}=R_{i j}-\frac{1}{2} g_{i j} R + \Lambda g_{i j},
\end{equation}
and $\mathscr{A}$ a gauge field depending on spatial coordinates and time.

\section{The Kerr Solution and the Lense-Thirring Precession}
\label{KERR}
Extended/Modified theories of gravity discussed in Sect. \ref{ETG} are very general and have no
predictive power in their fundamental  forms, due to the large number of degrees of freedom. In order to overcome this difficulty and to have  terms of comparison with GR, it is worth searching for  their  weak--field limit.  In this perspective,  it is possible to derive   expressions of phenomena, like the gyroscopic and Lense--Thirring precessions, to compare with experimental data. 

Before considering the PN and PPN formalism, let us recall how to get the Lense--Thirring precession starting from a generic Kerr spacetime. The  Hilbert--Einstein action 
\begin{equation}
S^{(GR)} = \int \sqrt{-g} \, R \, d^4 x,
\label{GRact}
\end{equation}
is the only gravitational action leading to analytic solutions for rotating compact objects, described by the Kerr metric. It is generally derived starting from  the line element 
\begin{eqnarray}
ds^2 &=& \mathcal{A}(t, r,\theta) dt^2 + \mathcal{B}(t, r,\theta) dr^{2} + \mathcal{C}(t, r,\theta) d\theta ^{2} \nonumber
\\
&+& \mathcal{D}(t, r,\theta) \sin^2 \theta d \varphi^2 + \mathcal{E}(t, r,\theta) dt\,d\varphi ,
\label{Kerr-int}
\end{eqnarray}
according to which a rotating body  exhibits a  frame-dragging precession. By plugging the interval  \eqref{Kerr-int} in the Einstein field equations derived from Action \eqref{GRact}, the rotating spherically symmetric solution in the equatorial plane reads
\begin{eqnarray}
ds^2 &=& \left( 1 - \frac{r_S}{r}\right) dt^2 - \frac{1}{1 - \frac{r_S}{r} +  \frac{J^2}{M^2 \, r^2}} dr^{2}- r^2 d\theta ^{2}  \nonumber
\\
&-&\left(r^2 + \frac{J^2}{M^2} + \frac{r_S J^2}{M^2 \, r} \right) d\varphi ^{2} -{\frac {2r_{S} \, J }{M r}}dt\,d\varphi\,,
\end{eqnarray}
where $J$ is the angular momentum of the rotating object with mass $M$ and $r_S$ the Schwarzschild radius $r_S = 2G_N M$. Note that setting $J=0$, the Schwarzschild solution occurs as a particular limit and the asymptotic flatness is recovered. An important effect related to rotating objects in GR is the so called \emph{frame dragging} (or equivalently \emph{Lense--Thirring effect}), according to which the spacetime metric is distorted by the rotation, giving rise to the precession of a test particle orbit. According to GR, the predicted value of the Lense-Thirring angular precession turns out to be
\begin{equation}
\Omega^{LT}_{(GR)} = \frac{r_S}{4 M r^3} \, J\,.
\end{equation}
It has been measured with high precision by Gravity Probe B experiment, which provided a precession of the order of  $\Omega^{LT} \sim 1.02 \cdot 10^{-4}$ arcseconds per day for the Earth \cite{Everitt:2011hp}. To better understand the physical meaning of the Lense--Thirring precession, let us consider  the spacetime dragging. In  Kerr-like metrics, the non-vanishing term $g_{03}$ yields new potentials in the weak--field limit. As the Newtonian potential $\Phi$  is defined by the second-order expansion of the time-component of the metric, namely
\begin{equation}
g_{00} = \frac{1}{1 - \frac{2 G_N M}{r}}\sim 1 + \frac{2G_N M}{r} = 1 + 2 \Phi,
\end{equation}
the other potentials arise from the expansion of the general metric \eqref{Kerr-int}. As a matter of fact,  spherically symmetric solutions in modified theories of gravity do not often occur analytically, especially in presence of matter. This is due to the fact that the extension/modification of the Hilbert--Einstein action yields higher--order field equations rarely  admitting exact solutions. 

The Schwarzschild spacetime is recovered when the angular momentum effects can be discarded. In this formalism, the metric $g_{\mu \nu}$ can be expanded by considering a perturbation $h_{\mu \nu}$ of the Minkowski spacetime, so that $h_{\mu \nu} \ll \eta_{\mu \nu}$ and
\begin{equation}
g_{\mu \nu} = \eta_{\mu \nu} + h_{\mu \nu}.
\end{equation}
The first-order Einstein tensor turns out to be
\begin{eqnarray}
G_{\mu \nu} &=& \frac {1}{2}(\partial _{\sigma }\partial _{\mu }h_{\nu }^{\sigma }+\partial _{\sigma }\partial _{\nu }h_{\mu }^{\sigma }-\partial _{\mu }\partial _{\nu }h-\square h_{\mu \nu } \nonumber
\\
&-&\eta _{\mu \nu }\partial _{\rho }\partial _{\lambda }h^{\rho \lambda }+\eta _{\mu \nu }\square h).
\end{eqnarray}
When higher--order corrections are considered, new scalar and vector potentials  arise in the approximation.  
Up to the fourth-order, a generic expansion of the metric tensor can be written as:
\begin{eqnarray}
g_{\mu \nu} &\sim & \left(\begin{array}{cc}1+g_{00}^{(2)}+g_{00}^{(4)}+\ldots & g_{0 i}^{(3)}+\ldots \\ g_{0 i}^{(3)}+\ldots & -\delta_{i j}+g_{i j}^{(2)}+\ldots\end{array}\right) \nonumber
\\
&=&\left(\begin{array}{cc}1+2 \Phi+2 \Xi & 2 A_{i} \\ 2 A_{i} & -\delta_{i j}+2 \Psi \delta_{i j}\end{array}\right).
\label{expansion}
\end{eqnarray}
It is worth noticing  that the second-order expansion of the element $g_{00}$, provides the Newtonian potential $\Phi$. Moreover, two other scalar potentials ($\Psi$ and $\Xi$) arise: the former comes from the second-order expansion of the metric tensor components $g_{ij}$ (in the Newtonian limit), while the latter is related to the fourth-order expansion of $g_{00}$ (the PN limit). Specifically, $\Phi(r)$, $\Psi(r)$ are scalar potentials proportional to $(v/c)^2$, $\Xi(r)$ is proportionals to $(v/c)^4$, while $A_i$ is a vector potential proportional to the power $(v/c)^3$. For the sake of clearness, in the metric \eqref{Kerr-int},  the elements $g_{00}$, $g_{0i}$ and $\delta^{ij} g_{ij}$ can be identified as
\begin{eqnarray}
&& g_{00} \equiv \mathcal{A}(t, r,\theta)
\\
&& g_{0i} = \mathcal{E}(t, r,\theta)
\\
&& g_{ij} \delta^{ij} = \mathcal{B}(t, r,\theta) + \mathcal{C}(t, r,\theta) + \mathcal{D}(t, r,\theta)
\end{eqnarray}
Most interestingly, the second-order expansion of the element $g_{0i}$, leads to a vector potential $A_i$. This is the potential linked to the rotations, whose curl operator magnitude (in analogy to the electromagnetic case) provides the Lense--Thirring precession. In GR, the angular precession can be derived analytically from the definition:
\begin{eqnarray}
\Omega^{LT}_{(GR)} &=& \frac{1}{2} (\epsilon^{ijk} \partial_i A_k) (\epsilon_{\ell n k} \partial^\ell A^k) \nonumber
\\
&=& \frac{G_N}{r^3} \sqrt{\left(\epsilon_{\ell k m} \partial^m \epsilon^{ijk} J_i x_j\right)^2} =  \frac{r_S}{4 M r^3} \, J,
\label{LTPR}
\end{eqnarray}
with obvious meaning of the symbols indicating the curl operator.
The shape of the vector potential $A_i$ can be found after replacing the potentials defined in the linearized field equations. However, the Kerr metric suffers the lack of  perfect fluids matching  boundary conditions between  external and  internal  solutions. In this context, extended/modified theories of gravity might solve the problem, containing free parameters which can be constrained by experimental observations. Reversing the argument, starting from a general unknown function of second order curvature invariants, it could be possible to constrain the form of the action by comparing the theoretically predicted value of the Lense--Thirring precession with the measured one. 

Notice that the only vector potential $\textbf{A}$, which provides corrections to GR, occurs when the second order invariant $R^{\mu \nu} R_{\mu \nu}$ is considered into the action. This is due to the fact that the scalar $R^{\mu \nu} R_{\mu \nu}$ is the only one which carries extra massive modes. However, the potential of  GR is recovered by imposing $f_Y \to 0$, namely $m_Y \to \infty$.

\subsection{The PPN formalism in higher--order--scalar--tensor gravity}
\label{STGRAV}
Let us  discuss now  the weak--field limit of  higher--order--scalar--tensor  gravity with the action \eqref{GENACT}. The aim is using the PN limit to find out the theoretical expression of the Lense--Thirring precession and, subsequently, to constrain the free parameters by the  experimental data provided by GINGER.

No exact solution, in a Kerr spacetime, has been found from the above action, but approximated solutions can be derived by replacing the metric \eqref{expansion} into the field Eqs.\eqref{FEGEN} considering the Taylor expansion \cite{Capozziello:2014mea}
\begin{eqnarray}
&& f\left(R, R_{\alpha \beta} R^{\alpha \beta}, \phi\right)= f_{R}\left(0,0, \phi^{(0)}\right) R \nonumber
\\
&& +\frac{f_{R R}\left(0,0, \phi^{(0)}\right)}{2} R^{2}+\frac{f_{\phi \phi}\left(0,0, \phi^{(0)}\right)}{2}\left(\phi-\phi^{(0)}\right)^{2} \nonumber
\\
&& +f_{R \phi}\left(0,0, \phi^{(0)}\right) R \phi+f_{Y}\left(0,0, \phi^{(0)}\right) R_{\alpha \beta} R^{\alpha \beta} , 
\end{eqnarray}

where $\phi^{(0)}$ is the zero-th order expansion of the scalar field $\phi$. From the linearization of the metric tensor, four potentials arise.  In order to find out  the effective form of the gyroscopic and Lense--Thirring precessions, only the scalar potential $\Phi(r)$ and the vector potential $A_i$ are needed. They read as:
\begin{eqnarray}
\Phi(r) &=& -\frac{G_N M}{r}\left[1+g(\xi, \eta) e^{-m_{R} \tilde{k}_{R} r} \right. \nonumber
\\
&+&\left. [1 / 3-g(\xi, \eta)] e^{-m_{R} \tilde{k}_{\phi} r} -\frac{4}{3} e^{-m_{Y} r}\right] 
\end{eqnarray}
and 
\begin{equation}
A_i = \frac{G_N}{r^{3}}\left[1-\left(1+m_{Y}r \right) e^{-m_{Y} r}\right] e_{ijk} r^j J^k,
\label{VPST}
\end{equation}
with the  parameters defined as
\begin{equation}
\begin{aligned} m_{R}^{2} &=-\frac{1}{3 f_{R R}\left(0,0, \phi^{(0)}\right)+2 f_{Y}\left(0,0, \phi^{(0)}\right)} \\ m_{Y}^{2} &=\frac{1}{f_{Y}\left(0,0, \phi^{(0)}\right)} \\ m_{\phi}^{2} &=-\frac{f_{\phi \phi}\left(0,0, \phi^{(0)}\right)}{2 \omega\left(\phi^{(0)}\right)} \\ \xi &=\frac{3 f_{R \phi}\left(0,0, \phi^{(0)}\right)^{2}}{2 \omega\left(\phi^{(0)}\right)} \\  \eta &=\frac{m_{\phi}}{m_{R}} \\ g(\xi, \eta) &=\frac{1-\eta^{2}+\xi+\sqrt{\eta^{4}+(\xi-1)^{2}-2 \eta^{2}(\xi+1)}}{6 \sqrt{\eta^{4}+(\xi-1)^{2}-2 \eta^{2}(\xi+1)}} \\ \tilde{k}_{R, \phi}^{2} &=\frac{1-\xi+\eta^{2} \pm \sqrt{\left(1-\xi+\eta^{2}\right)^{2}-4 \eta^{2}}}{2}. \end{aligned}
\end{equation}
The vector potential $A_i$ is written in terms of the angular momentum $J_i$.
Observations on  Solar System can provide the constraints on  the above  parameters. Following \cite{Capozziello:2014mea}, the gyroscopic precession is
\begin{eqnarray}
&& \Omega^{\mathrm{G}}_{(\mathrm{EG})}=-\left\{g(\xi, \eta)\left(m_{R} \tilde{k}_{R} r+1\right) F\left(m_{R} \tilde{k}_{R} \mathcal{R}\right) e^{-m_{R} \tilde{k}_{R} r} \right. \nonumber 
\\
&+& \frac{8}{3}\left(m_{Y} r+1\right) F\left(m_{Y} \mathcal{R}\right) e^{-m_{Y} r} + \left[\frac{1}{3}-g(\xi, \eta)\right]  \nonumber
\\
&\times&\left.\left(m_{R} \tilde{k}_{\phi} r+1\right) F\left(m_{R} \tilde{k}_{\phi} \mathcal{R}\right) e^{-m_{R} \tilde{k}_{\phi} r}\right\} \frac{\Omega^{\mathrm{G}}_{(\mathrm{GR})}}{3}, \nonumber
\end{eqnarray}
where $\mathcal{R}$ is the radius of the body, $\Omega^{\mathrm{G}}_{(\mathrm{EG})}$ is the extended gravity effect to be compared with
\begin{equation}
\Omega^G_{GR} = \left| \frac{3 G_N M}{2 r^3} \, \textbf{r} \times \textbf{v} \right|,
\label{OmegaGR}
\end{equation}
coming from GR. In Eq. \eqref{OmegaGR}, $\textbf{v}$ is the velocity of the Earth in the  considered location and 
\begin{equation}
F (m\tilde{k} \mathcal{R}) = 3\left[ \frac{(m \tilde{k}\mathcal{R} )\cosh (m \tilde{k}\mathcal{R}) - \sinh( m \tilde{k} \mathcal{R})}{m^3\tilde{k}^3 \mathcal{R}^3} \right] 
\end{equation}
is a geometric factor multiplying the Yukawa term determined by the parameters of the theory.  On the other hand, the Lense-Thirring precession  can be written in terms of  GR precession in Eq. \eqref{LTPR}. It reads
\begin{equation}
\Omega^{\mathrm{LT}}_{(\mathrm{EG})}=-e^{-m_{Y} r}\left(1+m_{Y} r+m_{Y}^{2} r^{2}\right) \Omega^{\mathrm{LT}}_{(\mathrm{GR})}\,,
\label{precLTST}
\end{equation}
where deviations with respect to GR are parameterized by $m_Y$, the effective mass related to the curvature invariant $R_{\mu\nu}R^{\mu\nu}$. In this way, the total Lense-Thirring precession can be recast as the sum:
\begin{eqnarray}
\Omega^{LT}_{TOT} &=& \Omega^{\mathrm{LT}}_{(\mathrm{EG})} + \Omega^{\mathrm{LT}}_{(\mathrm{GR})} \nonumber
\\
&=& \left[1 -e^{-m_{Y} r}\left(1+m_{Y} r+m_{Y}^{2} r^{2}\right)\right]\Omega^{\mathrm{LT}}_{(\mathrm{GR})},
\end{eqnarray}
from which
\begin{equation}
\left|\frac{\Omega^{LT}_{TOT}}{\Omega^{LT}_{GR}} - 1 \right| =e^{-m_{Y} r}\left(1+m_{Y} r+m_{Y}^{2} r^{2}\right),
\label{precLTST1}
\end{equation}
so that $\Omega^{\mathrm{LT}}_{(\mathrm{EG})} $ only quantifies the  deviations from $ \Omega^{\mathrm{LT}}_{(\mathrm{GR})} $.
It is worth noticing that Eq. \eqref{precLTST} can be obtained by applying the curl operator to the vector potential \eqref{VPST}, that is:
\begin{equation}
\textbf{A} = \frac{G_N}{r^3}\left[1 - \left( 1+ m_Y r  \right)e^{- m_Y r} \right]\textbf{r} \times \textbf{J}.
\end{equation}
Considering  data provided by Gravity Probe B satellite, orbiting at $h = 650$ km of altitude, the effective mass $m_Y$ is constrained to be
\begin{equation}
m_Y > 7.1 \cdot 10^{-5} m^{-1},
\end{equation}
given in terms of effective Compton  length.
This result can be obtained by considering, for the radial distance $r$, a sum between the Earth radius $R_{Earth}$ and the altitude $h$, namely $r =  R_{Earth} + h$. 

A further lower limit can be achieved by LARES satellite, whose observations on Lense--Thirring precession constrain the mass $m_Y$ to be
\begin{equation}
m_Y > 1.2 \cdot 10^{-6} m^{-1}.
\end{equation}
Considering Eq. \eqref{precLTST1}, our purpose is now  to further constrain the  parameter $m_Y$ by GINGER data for the gyroscopic and Lense--Thirring precessions. It is worth noticing that  GR, in the Earth frame,  predicts a gyroscopic precession  of the order  $\Omega^{G}_{GR} = 6606.1 \, mas/yr$ and a  Lense--Thirring, precession of the order  $\Omega^{LT}_{GR} = 37.2 \, mas/yr$. 

\begin{figure}
\begin{center}
\includegraphics[width=.50\textwidth]{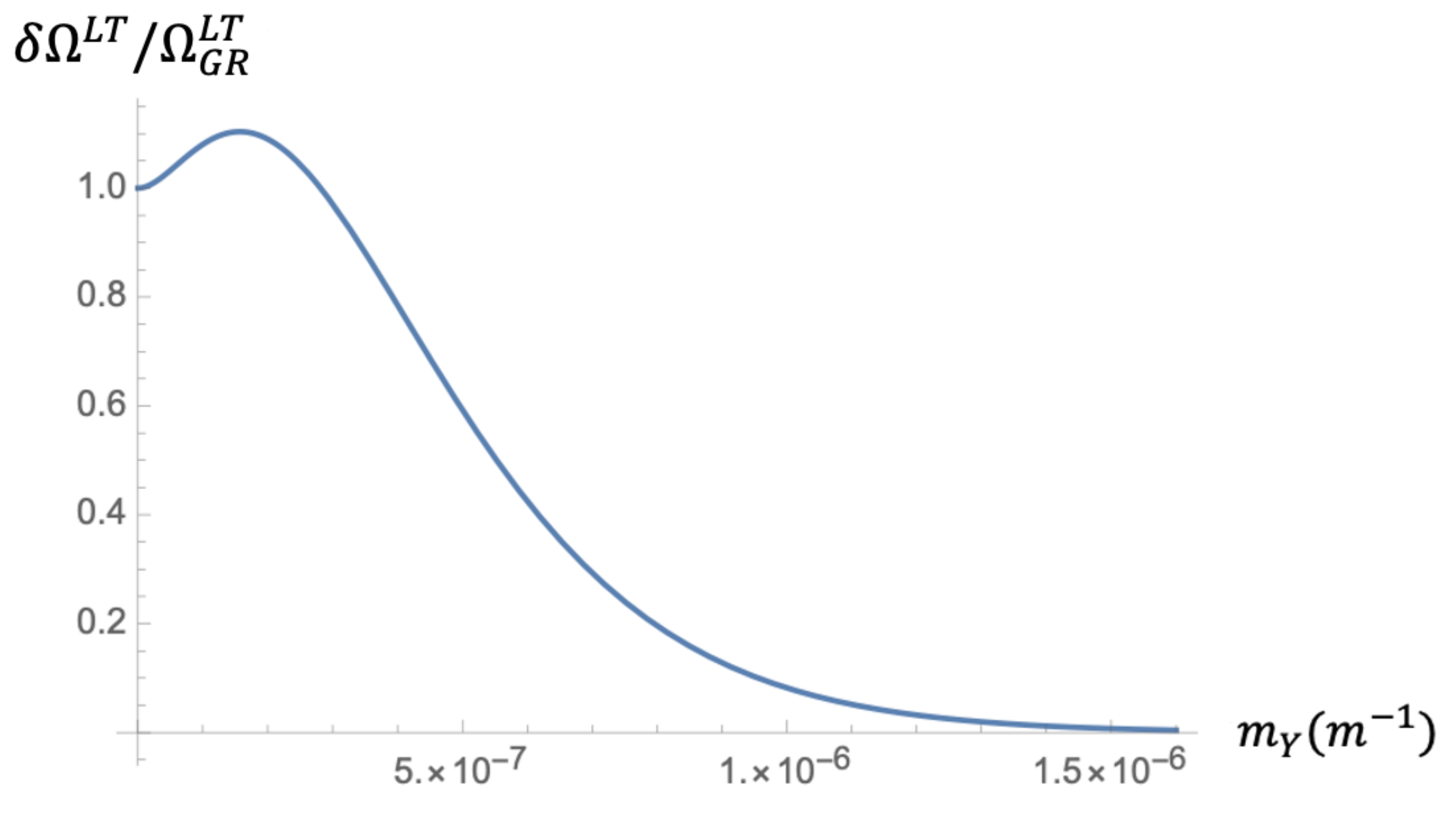}
\caption{$\delta \Omega^{LT}/\Omega^{LT}_{GR}$ as a function of $m_Y$. Note that GR is recovered in the limit $m_Y\rightarrow \infty$}
\label{fig:f1}
\end{center}
\end{figure}

From the relation \eqref{precLTST1},  it is possible to select a  range of compatibility for the effective mass $m_Y$ with GINGER data. To this purpose, we assume that the geodesic term (where GINGER can reach a precision of 1 part in $10^{4}$) and the Lense-Thirring term (where the expected precision is 1 part in $10^{3}$) can be disentangled. This can be done through different orientations of the ring lasers of the GINGER array. Replacing in Eq. \eqref{precLTST1} the distance between the point where the Lense--Thirring precession is measured and the center of Earth, the algebraic equation 
\begin{equation}
e^{-m_{Y} r}\left(1+m_{Y} r+m_{Y}^{2} r^{2}\right) \le 0.001,
\end{equation}
can be solved numerically providing
\begin{equation}
m_Y \ge 1.88 \cdot 10^{-6} m^{-1}.
\label{myGINGER}
\end{equation}
This result, obtained by an Earth-based experiment, put a further limit to  the mass $m_Y$ and, therefore, to higher-order scalar--tensor gravity models of the form $f(R, R^{\mu \nu} R_{\mu \nu}, \phi)$. The relation occurring between \\ $\delta \Omega^{LT}/\Omega^{LT}_{GR} \equiv (\Omega^{LT}_{TOT} - \Omega^{LT}_{GR})/\Omega^{LT}_{GR}$ and $m_Y$ is plotted in  Fig. \ref{fig:f1}.
It shows the high accuracy  needed to put physical  constraints on the effective mass $m_Y$. In particular, the more accurate the measure, the narrower the allowed range of $m_Y$ will be.

The value of $m_Y$ shown in Eq. \eqref{myGINGER} is comparable with the LARES one, but it can be considered model independent, since it is a 'local' measurement, which can be repeated at a different location, and does not require a gravity map of the Earth. 

\subsection{The PPN formalism in Horava--Lifshitz Gravity}
Let us discuss now the weak-field limit of Horava--Lifshitz gravity to find out an explicit form for the gyroscopic and Lense--Thirring precessions. As pointed out in \cite{Horava:2010zj, daSilva:2010bm}, in order  to be in agreement with   Solar System observations and  preserve the power-counting renormalizability at  ultraviolet scales, the matter Lagrangian must be chosen carefully. In \cite{Horava:2010zj, daSilva:2010bm},  the authors propose the following general matter Lagrangian
\begin{equation}
\Lagr_M =  \tilde{N} \sqrt{\tilde{g}} \mathcal{L}_{M}\left(\tilde{N}, \tilde{N}_{i}, \tilde{g}_{i j} ; \psi_{n}\right)
\end{equation}
where 
\begin{equation}
\begin{aligned} \tilde{N} &=\left(1-a_{1} \sigma\right) N \\ \tilde{N}^{i} &=N^{i}+N g^{i j} \nabla_{j} \phi \\ \tilde{g}_{i j} &=\left(1-a_{2} \sigma\right)^{2} g_{i j} \\ \sigma &=\frac{\mathscr{A}-\mathscr{B}}{N} \\ \mathscr{B} &=-\dot{\phi}+N^{i} \nabla_{i} \phi+\frac{1}{2} N \nabla^{i} \phi \nabla_{i}, \phi\end{aligned}
\end{equation}
with $a_1$, $a_2$ arbitrary constants. As previously pointed out, $\mathscr{A}$ is a gauge field depending on spatial coordinates and time, $N$ and $N^i$ are respectively Lapse and Shift functions, $g^{ij}$ is the three-dimensional metric. 
In the PPN formalism, the matter Lagrangian of this theory has a key role. Assuming a vanishing contribution of the cosmological constant $\Lambda = 0$, the gyroscopic and Lense--Thirring precessions in Horava--Lifshitz gravity turn out to be related with those of GR through the relations \cite{Radicella:2014jwa}:
\begin{equation}
\left|\frac{\Omega_{TOT}^{G} - \Omega^G_{GR}}{\Omega_{G R}^{G}}\right|=\left|\frac{2}{3}\left( \frac{G_{HL}}{G_N} a_{1}- \frac{a_{2}}{a_{1}} -1\right) \right|
\label{GYRHL}
\end{equation} 
and
\begin{equation}
\left|\frac{\Omega_{TOT}^{LT} - \Omega^{LT}_{GR}}{\Omega_{G R}^{LT}}\right|= \left| \frac{G_{HL}}{G_N} - 1\right|,
\label{G/GN}
\end{equation}
where the standard Newton constant $G_N$ must be distinguished from the effective gravitational constant of Horava--Lifshitz gravity $G_{HL}$. In this way the measure on the Lense--Thirring precession can be used to constrain the effective gravitational constant, which, in turn, can be replaced in Eq. \eqref{GYRHL} to find a graphical relation between $a_1$ and $a_2$. 

\begin{figure}
\begin{center}
\centering
\includegraphics[width=.35\textwidth]{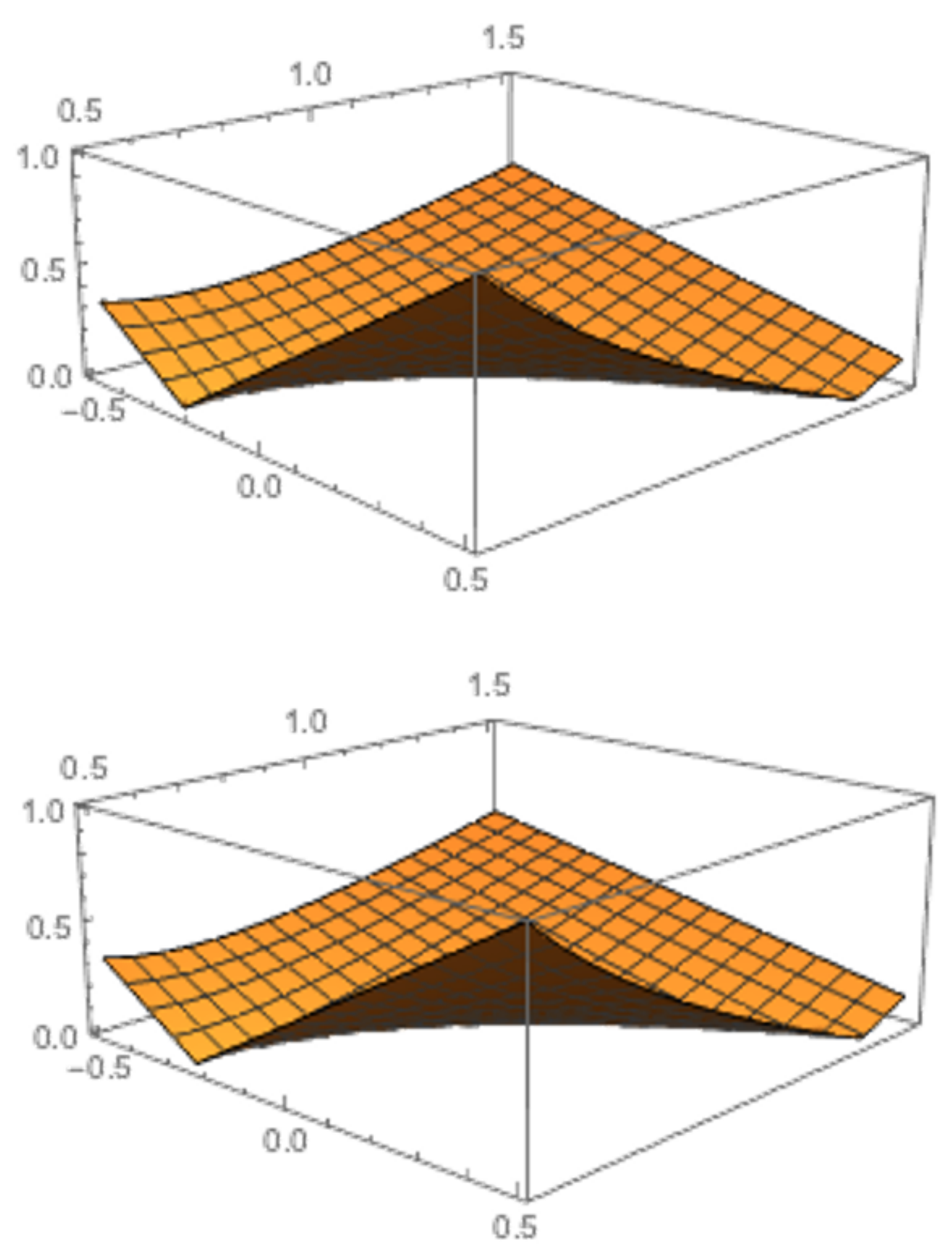}
\caption{$\delta \Omega^G/\Omega^G_{GR}$ as a function of $a_1$ and $a_2$, with $G_{HL}/G_N$ fixed to $0.999$ (top) and 1.001 (bottom).} 
\label{fig:f2}
\end{center}
\end{figure}

\begin{figure}
\begin{center}
\centering
\includegraphics[width=.48\textwidth]{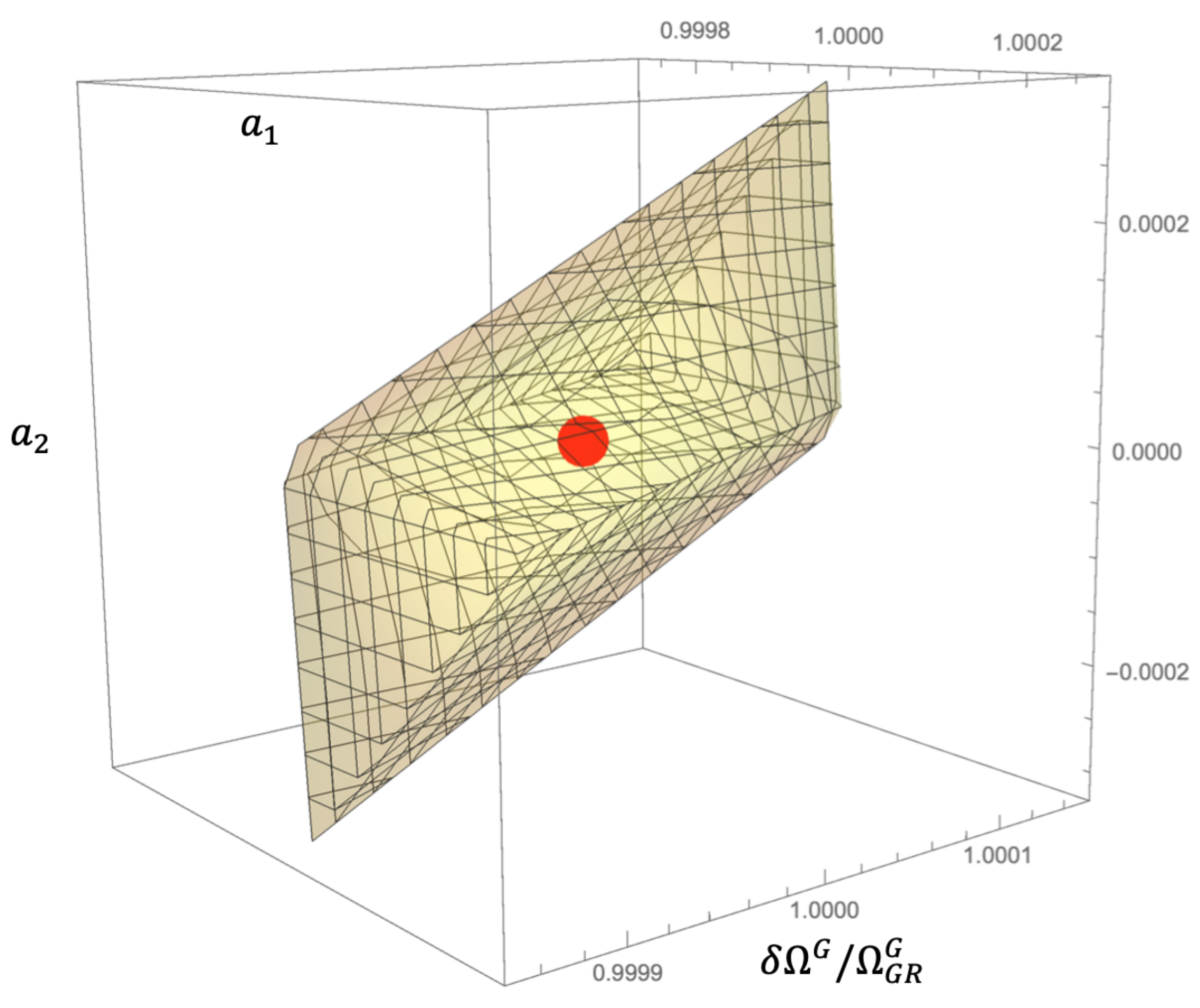}
\caption{Graphical representation of Eq. \eqref{diseq}. Red dot denotes the values of free parameters for which GR is recovered.}
\label{fig:f3}
\end{center}
\end{figure}

In order to adapt the GINGER measures to the Horava--Lifshitz gravity, in the weak-field limit, we consider the  frequency difference of light for two beams circulating in a laser cavity in opposite directions. This can be recast as a time difference between the right-handed beam propagation time and the left-handed one, namely
\begin{equation}
\delta \tau=P \lambda \Omega_{S},
\end{equation}
with 
\begin{equation}
\Omega_{S}=-\frac{2 \sqrt{g_{00}}}{P \lambda} \oint \frac{g_{0 i}}{g_{00}} d s^{i}.
\label{taud}
\end{equation}
In the above equations,  $P$ is the perimeter of the ring, $\lambda$ the laser wavelength, and $\Omega_S$ the splitting in terms of frequency between the two beams. Replacing the form of $g_{00}$ and $g_{0i}$ for Horava--Lifshitz gravity in Eq. \eqref{taud}, $\Omega_S$ turns out to be  \cite{Radicella:2014jwa}
\begin{eqnarray}
\Omega_{S}&=&\frac{4 A}{P \lambda} \Omega_{E}\left[\cos (\theta+\alpha) \right. \nonumber
\\
&-& \left(1+\frac{G_{HL}}{G_{N}} a_{1}-\frac{a_{2}}{a_{1}}\right) \frac{G_{HL} M}{R} \sin \alpha \sin \theta \nonumber
\\
 &-& \left. \frac{G_{HL} I_{E}}{R^{3}}(2 \cos \theta \cos \alpha+\sin \theta \sin \alpha)\right], 
\label{OMEGAS}
\end{eqnarray}
with $A$ being the area encircled by the light beams, $\alpha$ the angle between the local radial direction and the normal to the plane of the array-laser ring, $\theta$ the colatitude of the laboratory, $\Omega_E$ the rotation rate of the Earth as measured in the local reference frame and $I_E$ the Earth momentum of inertia. The first term in the RHS is the Sagnac term, the second is the gyroscopic term, while last is the Lense--Thirring term. Notice that by setting $G_{HL} = G_N$, the splitting in terms of frequency reduces to that of GR, as expected. The presence of two rings yields a dynamic measure of the angle, so that the overall precision of GINGER is 1/100 in the Lense--Thirring term and 1/1000 in the geodetic term. From Eq. \eqref{G/GN}, we obtain that the effective gravitational constant of Horava--Lifshitz gravity is related to the Newtonian constant through the numerical relation
\begin{equation}
0.999 \, G_N <G_{HL} < 1.001 \, G_N
\end{equation}
Setting $\displaystyle \frac{G_{HL}}{G_N} = 0.999$ and replacing  into Eq. \eqref{OMEGAS}, it turns out that the coupling constants $a_1$ and $a_2$ satisfy the relation
\begin{eqnarray}
&&  a_1 (0.999 a_1 - 0.99985 ) < a_2 <  a_1 (0.999 a_1-1.00015 ) \nonumber
\\
&& \text{if} \,\,\,  a_1<0, \nonumber
 \\ \nonumber \\
&&  a_1 (0.999 a_1-1.00015 )< a_2 <  a_1 (0.999 a_1 - 0.99985 )  \nonumber
\\
&& \text{if} \,\,\,  a_1>0.
\label{a12}
\end{eqnarray}
The graphical representation is shown in Fig. \ref{fig:f2}. 
Similarly, by setting $\displaystyle \frac{G_{HL}}{G_N} = 1.001$, we obtain
\begin{eqnarray}
&&  a_1 (1.001 a_1 - 0.99985 ) < a_2 <  a_1 (1.001 a_1-1.00015 ) \nonumber
\\
&&\text{if} \,\,\,  a_1<0 \nonumber
 \\\nonumber \\
&&  a_1 (1.001 a_1-1.00015 )< a_2 <  a_1 (1.001 a_1 - 0.99985 ) \nonumber
\\
&& \text{if} \,\,\,  a_1>0.
\label{a12_1}
\end{eqnarray}
The graphical representation is also reported in Fig \ref{fig:f2}. 
The ranges \eqref{a12} and \eqref{a12_1} can be obtained by considering the precision of GINGER in the geodesic term, that is expected to be 1 part in $10^4$. Moreover we only plotted ranges of $a_1$ and $a_2$ which can provide small corrections to GR. Indeed, we considered $a_1 \in [0.5,1.5]$ and $a_1 \in [-0.5, 0.5]$, so that GR is recovered as soon as  $a_1 = 1$ and $a_2 = 0$.  In what follows, we let $G_{HL}/G_N$ varies between $0.999$ and $1.001$, which are the validity limits provided by the analysis of the Lense-Thirring term. Therefore, the relations between $a_1$, $a_2$ and $G_{HL}/G_N$, that is
\begin{equation}
\left|\frac{\Omega_{TOT}^{G} - \Omega^G_{GR}}{\Omega_{G R}^{G}}\right|=\left|\frac{2}{3}\left( \frac{G_{HL}}{G_N} a_{1}- \frac{a_{2}}{a_{1}} -1\right) \right| < 10^{-4}
\label{diseq}
\end{equation}
is plotted in Fig. \ref{fig:f3},
which shows numerical constraints to the matter Lagrangian of Horava--Lifshitz gravity. In order to uniquely find $a_1$ and $a_2$, a further analysis is needed. We aim to use GINGER data in future works, with the purpose to ameliorate the precision of the numerical values of the free parameters.

\section{Discussion and Conclusions}
\label{CONCL}
We considered two modified theories of gravity and constrained the corresponding free parameters by GINGER experimental data. The former theory is based on  extensions of the Hilbert--Einstein action, and belongs to the so-called  Extended Theories of Gravity. Extending GR leads to higher--order field equations  carrying   for the gravitational field.  From the weak-field limit of the theory,  it is possible to relate the gyroscopic and Lense--Thirring precessions with the parameters of these further degrees of freedom. GINGER experimental data of these precessions put an upper limit to the first derivative of the function $f(R, R^{\mu \nu} R_{\mu \nu}, \phi)$ with respect to the second-order curvature invariant $R^{\mu \nu} R_{\mu \nu}$. Specifically we get
\begin{equation}
m_Y \equiv \sqrt{\frac{1}{f_{Y}\left(0,0, \phi^{(0)}\right)}} \ge 1.88 \cdot 10^{-6} m^{-1}.
\end{equation} 
If compared to the results provided by Lares and GP-B, we notice that GINGER can provide a stronger constraint to the effective mass with respect to the latter. 

The second theory here discussed belongs to the class of alternative theories of gravity aimed to relax some assumptions of GR and to construct actions which better fit the high-energy shortcomings suffered by Einstein gravity. We considered a specific matter Lagrangian of Horava-Lifshitz gravity, which preserves the unitarity and allows to fit the Solar System observations. Such Lagrangian, in the most general form, introduces in the theory two arbitrary parameters $a_1$ and $a_2$, which cannot be theoretically constrained. We used the weak--field limit to find a relation between the two constants and the value of the effective gravitational constant in Horava--Lifshitz gravity. In particular, from the Lense-Thirring precession we obtain
\begin{equation}
0.999 \, G_N <G_{HL} < 1.001 \, G_N.
\end{equation} 
By means of this result, it turns out that $a_1$ and $a_2$ can be fixed through data on Lense--Thirring precession, according to which relations \eqref{a12} and \eqref{a12_1} hold. Moreover, considering a precision of 1/10000 for the geodesic term, Eq. \eqref{diseq} provides a numerical relation between $a_1$, $a_2$ and $G_{HL}/G_N$, plotted in Fig. \ref{fig:f3}.

The analysis of this work represents a first step towards constraining modified theories of gravity and, in general, metric theories through GINGER data. 

Some points need to be stressed in conclusion. With respect to results coming from satellites \cite{Capozziello:2014mea}, GINGER measurements avoid issues related  to the dynamical configurations of space-based experiments. For example, thanks to the experimental set up, errors and noise coming from considering fine  gravity maps can be avoided. Finally, being the experiment essentially static,  problems related to the timing and the synchronization between different reference frames (e.g. Earth and satellite) can be removed.  

As final remark, we can say that relatively simple experiments like GINGER can achieve and ameliorate the performances of space-based setups in measuring gravitational parameters. This could be the starting point for a systematic investigation of theories of gravity by Earth-based experiments.

\section*{Acknowledgments}
The authors acknowledge the support of  {\it Istituto Nazionale di Fisica Nucleare} (INFN) ({\it iniziative specifiche} MOONLIGHT2 and  GINGER).

\twocolumn[
  \begin{@twocolumnfalse}
\resizebox{16.6cm}{!}{
\begin{centering}
\begin{tabular}{c|c|cl}
\hline
  Modified  Gravity Model  & Corrected potential & Yukawa parameters \\
\hline &&&
\\
 $f(R)$ & $\begin{array}{ll} \Phi(r)=-\frac{G_N M}{r}\biggl[1+\alpha\,e^{-m_R r}\biggl] \\ \displaystyle \textbf{A} = \frac{G_N}{r^3} \, \textbf{r} \times \textbf{J} \end{array}$ & $\begin{array}{ll}{m_R}^2\,=\,-\frac{f_R(0)}{6f_{RR}(0)}\end{array}$ \\ \\
\hline
 & & &\\ 
$f(R,\Box R)= R+ a_0R^2+a_1R\Box R$ & $\begin{array}{ll} \Phi(r)=-\frac{G_NM}{r}\left(1+c_0e^{(-r/l_0)}+c_1e^{(-r/l_1)}\right) \\ \displaystyle \textbf{A} = \frac{G_N}{r^3} \, \textbf{r} \times \textbf{J} \end{array}$ & $\begin{array}{ll}{c_{0,1}}\,=\,\frac{1}{6}\mp\frac{a_0}{2\sqrt{9a_0^2+6a_1}}\\\\  l_{0,1}\,=\,\sqrt{-3a_0\pm\sqrt{9a_0^2+6a_1}}\end{array}$ \\
\hline &&&
\\
$f(R,\Box R,..\Box^k R)= R+\Sigma_{k=0}^p a_kR\Box^k R$ & $\begin{array}{ll} \Phi(r)=-\frac{G_NM}{r}\left(1+\Sigma_{k=0}^{p}c_i\exp(-r/l_i)\right) \\ \displaystyle \textbf{A} = \frac{G_N}{r^3} \, \textbf{r} \times \textbf{J} \end{array}$ & $c_i\,,\;l_i$ are functions of $a_k$. See \cite{Quandt:1990gc}. \\ \\
\hline
 & & &\\
   $f(R,\,R_{\alpha\beta}R^{\alpha\beta})$ & $\begin{array}{ll}\Phi(r)=-\frac{G_NM}{r}\biggl[1+\frac{1}{3}\,e^{-m_R r}-\frac{4}{3}\,e^{-m_Y r}\biggl] \\ \displaystyle \textbf{A} = \frac{G_N}{r^3} \, \textbf{r} \times \textbf{J} \end{array}$ & $\begin{array}{ll}{m_R}^2\,=\,-\frac{1}{3f_{RR}(0)+2f_Y(0)} \\\\{m_Y}^2\,=\,\frac{1}{f_Y(0)}\,,\, Y=R_{\mu\nu}R^{\mu\nu}
\end{array}$ \\ \\
\hline  
& & & \\  
$f(R,\,\cal{G})$\,, \; ${\cal G}= R^2-4R_{\mu\nu}R^{\mu\nu}+ R_{\alpha\beta\mu\nu}R^{\alpha\beta\mu\nu}$& $\begin{array}{ll}\Phi(r)=-\frac{G_NM}{r}\biggl[1+\frac{1}{3}\,e^{-m_1 r}-\frac{4}{3}\,e^{-m_2 r}\biggl] \\ \displaystyle \textbf{A} = \frac{G_N}{r^3} \, \textbf{r} \times \textbf{J} \end{array}$  & $\begin{array}{ll}{m_1}^2\,=\,-\frac{1}{3f_{RR}(0)+2f_Y(0)+2f_Z(0)} 
\\\\{m_2}^2\,=\,\frac{1}{f_Y(0)+4f_Z(0)}\,,\,Z=R_{\alpha\beta\mu\nu}R^{\alpha\beta\mu\nu}
\end{array}$\\ \\
\hline  & & & 
\\  
$f(R,\,\phi)+\omega(\phi)\phi_{;\alpha}\phi^{;\alpha}$ &
$\begin{array}{ll}\Phi(r)=-\frac{G_NM}{r}\biggl[1+g(\xi,\eta)\,e^{-m_R\tilde{k}_R r}+\\\\\qquad\qquad+[1/3-g(\xi,\eta)]\,e^{-m_R\tilde{k}_\phi r}\biggr] \\ \\  \displaystyle \textbf{A} = \frac{G_N}{r^3} \, \textbf{r} \times \textbf{J}\end{array}$
&
$\begin{array}{ll}{m_R}^2\,=\,-\frac{1}{3f_{RR}(0,\phi^{(0)})}\\\\{m_\phi}^2\,=\,-\frac{f_{\phi\phi}(0,\phi^{(0)})}{2\omega(\phi^{(0)})}\\\\\xi\,=\,\frac{3{f_{R\phi}(0,\phi^{(0)})}^2}{2\omega(\phi^{(0)})}\\\\\eta\,=\,\frac{m_\phi}{m_R}\\\\g(\xi,\,\eta)\,=\,\frac{1-\eta^2+\xi+\sqrt{\eta^4+(\xi-1)^2-2\eta^2(\xi+1)}}{6\sqrt{\eta^4+(\xi-1)^2-2\eta^2(\xi+1)}}\\\\{\tilde{k}_{R,\phi}}^2\,=\,\frac{1-\xi+\eta^2\pm\sqrt{(1-\xi+\eta^2)^2-4\eta^2}}{2}
\end{array}$ \\
\hline & & & 
\\  
$f(R,\,R_{\alpha\beta}R^{\alpha\beta},\phi)+\omega(\phi)\phi_{;\alpha}\phi^{;\alpha}$
&
$\begin{array}{ll}\Phi(r)=-\frac{G_NM}{r}\biggl[1+g(\xi,\eta)\,e^{-m_R\tilde{k}_R\,r}+\\\\\,\,\,\,+[1/3-g(\xi,\eta)]\,e^{-m_R\tilde{k}_\phi\,r}-\frac{4}{3}\,e^{-m_Y r}\biggr] \\ \\ \\  \\ \displaystyle \textbf{A} = \frac{G_N}{r^3}\left[1 - \left( 1+ m_Y r  \right)e^{- m_Y r} \right]\textbf{r} \times \textbf{J}\end{array}$ 
&
$\begin{array}{ll}{m_R}^2\,=\,-\frac{1}{3f_{RR}(0,0,\phi^{(0)})+2f_Y(0,0,\phi^{(0)})}\\\\{m_Y}^2\,=\,\frac{1}{f_Y(0,0,\phi^{(0)})}\\\\{m_\phi}^2\,=\,-\frac{f_{\phi\phi}(0,0,\phi^{(0)})}{2\omega(\phi^{(0)})}\\\\\xi\,=\,\frac{3{f_{R\phi}(0,0,\phi^{(0)})}^2}{2\omega(\phi^{(0)})}\\\\\eta\,=\,\frac{m_\phi}{m_R}\\\\g(\xi,\,\eta)\,=\,\frac{1-\eta^2+\xi+\sqrt{\eta^4+(\xi-1)^2-2\eta^2(\xi+1)}}{6\sqrt{\eta^4+(\xi-1)^2-2\eta^2(\xi+1)}}\\\\{\tilde{k}_{R,\phi}}^2\,=\,\frac{1-\xi+\eta^2\pm\sqrt{(1-\xi+\eta^2)^2-4\eta^2}}{2}
\end{array}$ \\
\hline 
\end{tabular} 
\end{centering} }
\\ Table I: Yukawa-like corrections are a general feature of several modified gravity models. In particular,  they emerge in Extended Theories of Gravity which are natural extension of General Relativity \cite{Capozziello:2011et}. In some sense, further degrees of freedom, related to  higher-order terms or scalar fields, give rise to these corrections in the weak field limit. This is a general result as discussed in  \cite{Quandt:1990gc}. In the Table, we report examples of modified gravity models showing  Yukawa-like corrections in the post-Newtonian limit. Detailed discussions of these results   are reported in \cite{Nojiri:2005jg, Quandt:1990gc, Capozziello:2014mea, DeLaurentis:2013ska, Stabile:2010mz}. 
  \end{@twocolumnfalse}
]

\end{document}